\documentclass{emulateapj}

\begin{document}
\slugcomment{Accepted by AJ}

\title{The Century Survey Galactic Halo Project II:  Global Properties 
and the Luminosity Function of Field Blue Horizontal Branch Stars}

\author{Warren R.\ Brown,
	Margaret J.\ Geller,
	Scott J.\ Kenyon,
	Michael J.\ Kurtz}

\affil{Smithsonian Astrophysical Observatory, Harvard-Smithsonian Center 
for Astrophysics, 60 Garden St, Cambridge, MA 02138}

\author{Carlos Allende Prieto}

\affil{McDonald Observatory and Department of Astronomy, 
University of Texas, Austin, TX 78712}

\author{Timothy C.\ Beers}

\affil{Department of Physics \& Astronomy and JINA: Joint Institute 
for Nuclear Astrophysics, Michigan State University,
East Lansing, MI 48824}

\and

\author{Ronald Wilhelm}

\affil{Department of Physics, Texas Tech University, Lubbock, TX 79409}

\shorttitle{Century Survey Galactic Halo Project II}
\shortauthors{W.\ R.\ Brown et al.}

% \clearpage
\begin{abstract}
	We discuss a 175 deg$^2$ spectroscopic survey for blue horizontal
branch (BHB) stars in the Galactic halo. We use the Two Micron All Sky
Survey (2MASS) and the Sloan Digital Sky Survey (SDSS) to select BHB
candidates, and find that the 2MASS and SDSS color-selection is 38\% and
50\% efficient, respectively, for BHB stars. Our samples include one
likely run-away B7 star 6 kpc below the Galactic plane. The global
properties of the BHB samples are consistent with membership in the halo
population: the median metallicity is [Fe/H]=$-1.7$, the velocity
dispersion is 108 km s$^{-1}$, and the mean Galactic rotation of the BHB
stars $3<|z|<15$ kpc is $-4\pm30$ km s$^{-1}$.  We discuss the theoretical
basis of the Preston, Shectman \& Beers $M_V$-color relation for BHB
stars, and conclude that intrinsic shape of the BHB $M_V$-color relation
results from the physics of stars on the horizontal branch.  We calculate
the luminosity function for the field BHB star samples using the
Efstathiou, Ellis, \& Peterson maximum-likelihood method which is unbiased
by density variations. The field BHB luminosity function exhibits a steep
rise at bright luminosities, a peak between $0.8 < M_V < 1.0$, and a tail
at faint luminosities. We compare the field BHB luminosity functions with
the luminosity functions derived from sixteen different globular cluster
BHBs. Kolmogorov-Smirnov tests suggest that field BHB stars and BHB stars
in globular clusters share a common distribution of luminosities, with the
exception of globular clusters with extended BHBs.

\end{abstract}


% \clearpage
\section{INTRODUCTION}

        Mapping the stellar halo requires objects that are sufficiently
luminous to observe at large distances, yet common enough to sample the
halo densely. In \citet{brown03}, hereafter Paper I, we introduced the
Century Survey Galactic Halo Project, a photometric and spectroscopic
survey from which we selected blue horizontal branch (BHB) stars as probes
of the Milky Way halo. BHB stars meet our criteria for tracer samples:  
they are intrinsically luminous and are quite numerous, with a number
density in the halo that exceeds that of RR Lyraes by roughly a factor of
ten \citep{preston91}. The spectral types of BHB stars are typically
around A0, bluer than most competing stellar populations. As a result,
candidate BHB stars in the halo are relatively easy to select by broadband
colors alone.

	In Paper I we described the detailed stellar spectral analysis
techniques developed for the Century Survey Galactic Halo Project. In this
paper we investigate the mean Galactic rotation, metallicity, and
luminosity function of the halo BHB stars in the context of a
complementary 175 deg$^2$ spectroscopic survey. This new survey extends
the work of the original Century Survey Galactic Halo Project by making
use of two large-area, multi-passband imaging surveys: (1) the Two Micron
All Sky Survey \citep[2MASS;][]{cutri03} and (2) the Sloan Digital Sky
Survey \citep[SDSS;][]{york00}.

	Previous spectroscopic surveys of field BHB stars have identified
BHB stars over large (several 10$^3$ deg$^2$) areas of sky to shallower
depths \citep{pier83,wilhelm99b}, or over small ($\sim$10$^2$ deg$^2$)
areas of sky to greater depths
\citep{sommer89,arnold92,kinman94,clewley04,kinman05} than the Century
Survey Galactic Halo Project.  The exception is the recently published
sample of 1170 BHB stars observed by the SDSS as mis-identified quasars or
as filler objects in low density regions \citep{sirko04a,sirko04b}.  In
comparison, our spectroscopic survey of BHB stars is cleanly selected and
100\% complete within our color and magnitude selection limits.  Combined
with the original Century Survey sample, we have 157
spectroscopically-identified BHB stars over 239 deg$^2$ of sky.

	In \S 2 we describe the sample selection and spectroscopic
observations of the new 175 deg$^2$ region and discuss selection
efficiencies for BHB stars.  In \S 3 we discuss the basis of BHB
luminosity-color-metallicity relations, and analyze the global kinematic
and abundance properties of our BHB samples.  In \S 4 we calculate the
luminosity functions for our field BHB star samples, and compare them with
luminosity functions derived from globular cluster data.  We summarize our
results and conclude in \S 5.

\section{SAMPLE SELECTION}

\subsection{Selection Region}

	The original Century Survey Galactic Halo Project contains BHB
stars selected by $(V-R)_0$ or $(J-H)_0$ colors.  Here we make use only 
of the sample selected with $(V-R)_0<0.3$, the ``Century Survey'' sample
\citep{brown03}.  The original Century Survey sample covers a
$1^{\circ}\times64^{\circ}$ slice located $8\fh5 < \alpha_{1950} < 13\fh5,
29^\circ < \delta_{1950} < 30^\circ$ and contains 39
spectroscopically-confirmed BHB stars in the magnitude range $13<V_0<16.5$
mag.

\begin{figure}
 \includegraphics[width=3.35in]{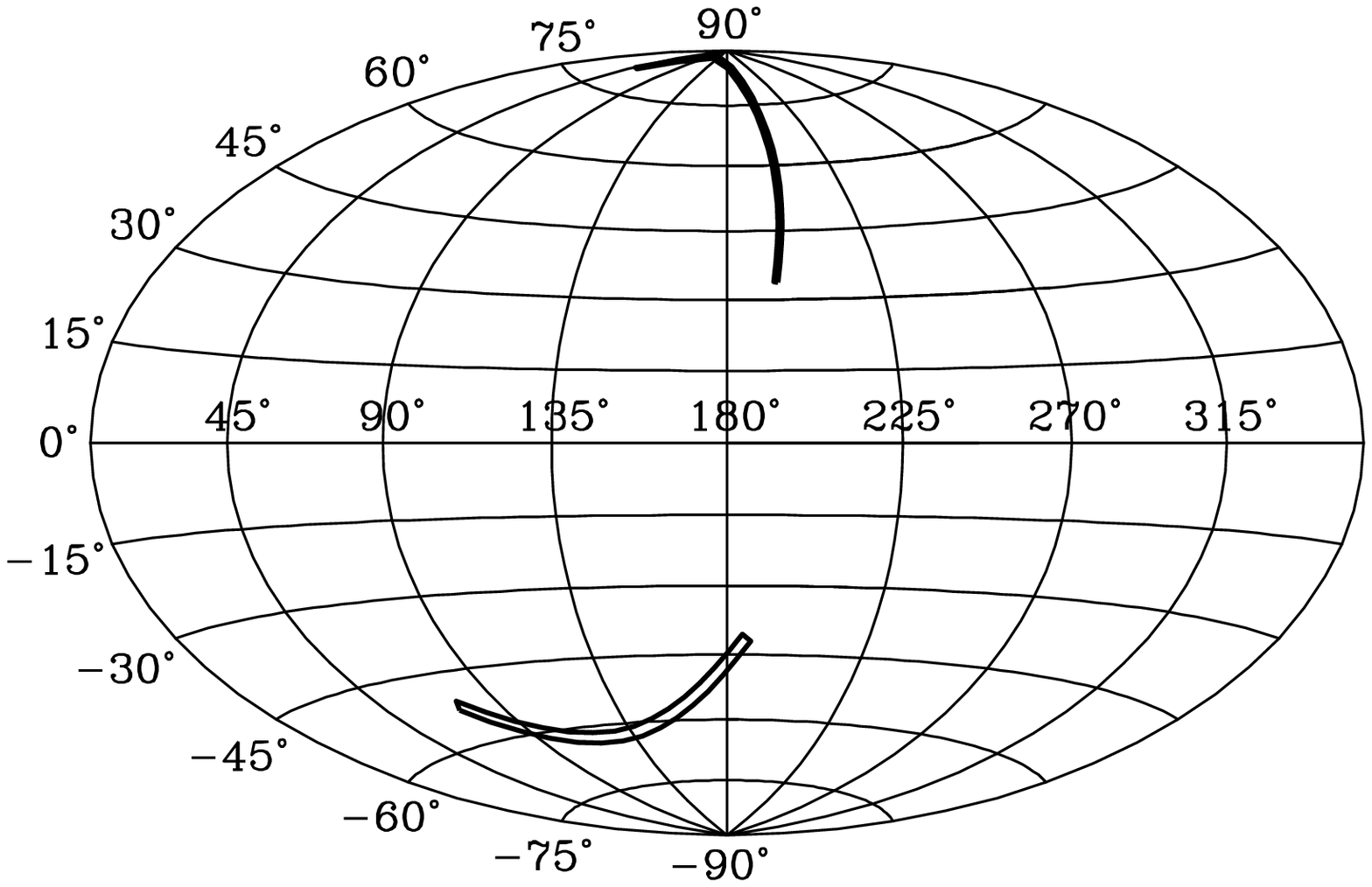}
 \caption{ \label{fig:sky}
	Sky map in Galactic coordinates.  The Century Survey slice $8\fh5
< {\rm RA_{B1950}} < 13\fh5$, $29\fdg0 < {\rm Dec_{B1950}} < 30\fdg0$ is
located in the north Galactic hemisphere.  Our new survey slice $23\fh0 <
{\rm RA_{J2000}} < 3\fh67$, $-1\fdg25^{\circ} < {\rm Dec_{J2000}} <
+1\fdg25$ is located in the south Galactic hemisphere.}
\end{figure}

	Here we select BHB candidate stars from the 2MASS and SDSS surveys
in a complementary region located along the celestial equator $23^{\rm
h}0^{\rm m}0^{\rm s} < \alpha_{J2000} < 3^{\rm h}40^{\rm m}0^{\rm s}$,
$-1^{\circ}15'0''< \delta_{J2000} < +1^{\circ}15'0''$.  Fig.\
\ref{fig:sky} is a plot of this $70^{\circ} \times2.5^{\circ}$ region in
Galactic coordinates.  The survey is located predominantly at
$b<-45^{\circ}$, in a region that cleanly samples the halo in the
\citet{brown04} BHB-candidate maps.

\subsection{2MASS Selection}

	The 2MASS catalog provides uniform $JHK$ photometry over the
entire sky.  In \citet{brown04}, we matched the original Century Survey
sample to 2MASS and showed that 2MASS colors select A-type stars with
$\sim$80\% efficiency.  The A-type stars are all good BHB candidates in
our high Galactic latitude survey region.

        We have selected 90 BHB candidates from the 2MASS catalog in the
magnitude range $12.5 < J_0 < 15.5$; BHB candidates have colors in the
ranges $-0.2 < (J-H)_0 < 0.1$ and $-0.1 < (H-K)_0 < 0.1$, following
\citet{brown04}.  Our upper color limits result in a high selection
efficiency but a reduced completeness for BHB stars.  
Comparison with the original Century Survey sample shows that
our color selection samples 65\% of the BHB population \citep{brown04}.

\begin{figure}
 \centerline{\includegraphics[width=3.0in]{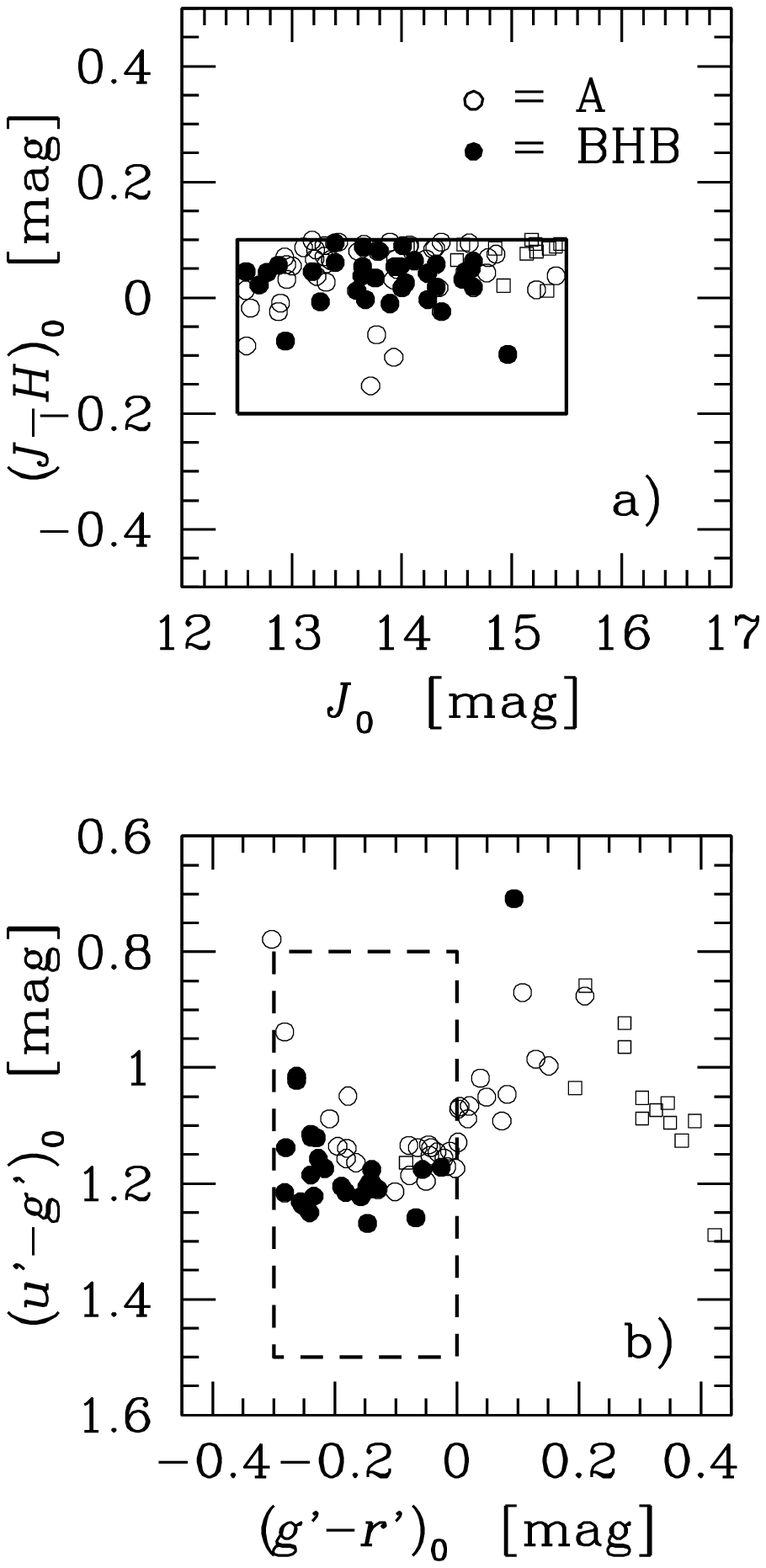}}
 \caption{ \label{fig:phot2m}
        The 2MASS-selected BHB candidate sample.  Panel (a) shows the
distribution of $(J-H)_0$ and $J_0$; the solid box shows the sample
selection region.  Panel (b) shows the distribution of $(u'-g')_0$ and
$(g'-r')_0$ colors; the dashed box shows the SDSS-sample selection region
for comparison. }
\end{figure}

	It is important to note that we have selected objects using
de-reddened colors and magnitudes, using extinctions from
\citet{schlegel98}.  The surface density of the 2MASS-selected BHB
candidates is 0.5 deg$^{-2}$.

	We have matched up our 2MASS-selected BHB candidates with the
publicly available SDSS data:  SDSS photometry presently exists for 65 of
the 90 objects.  Approximately half of the matched objects have SDSS
colors consistent with early A-type stars; the remainder follow the
stellar locus to F-type stars (see Fig.\ \ref{fig:phot2m}).

\subsection{SDSS Selection}

\begin{figure}
 \centerline{\includegraphics[width=3.0in]{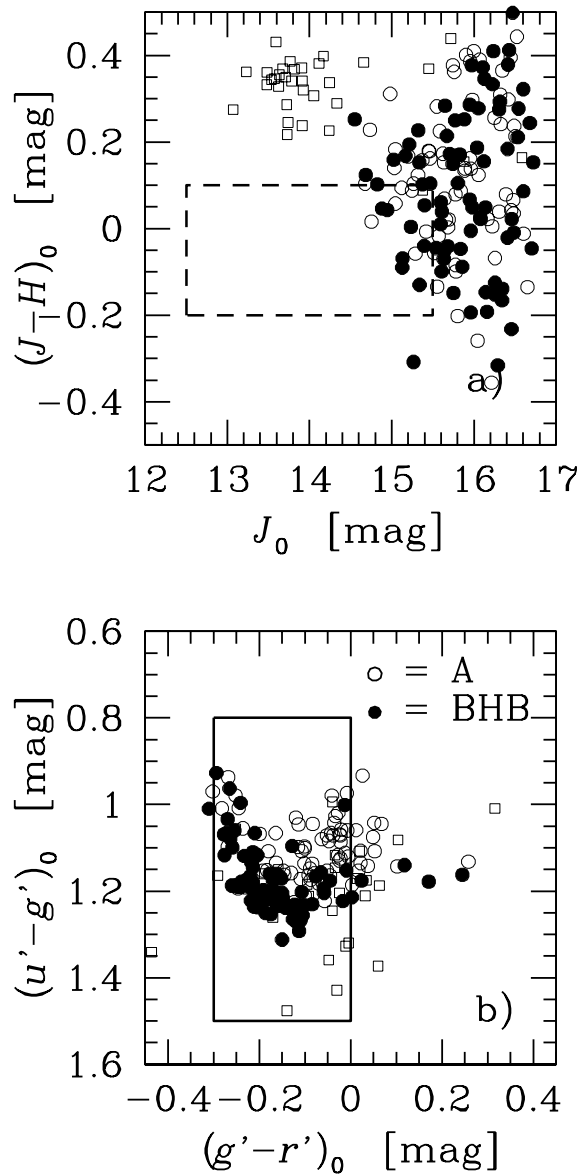}}
 \caption{ \label{fig:photss}
        Same as Figure \ref{fig:phot2m}, but for the SDSS-selected BHB
candidate sample.  Objects outside the SDSS-sample selection region (solid
box) are BHB candidates originally selected by ``model'' magnitudes from
the Early Data Release; here we plot Data Release 1 Petrosian magnitudes
that we find are better behaved at bright magnitudes.}
\end{figure}

	The SDSS has released five passband photometry for limited areas
of the sky that can be used to select A-type stars efficiently.
	We selected 194 BHB candidates in the magnitude range $15 < g'_0 <
17$ from the SDSS Early Data Release \citep{stoughton02} and Data Release
1 \citep{abazajian03}.  We follow \citet{yanny00} and select BHB
candidates with $-0.3 < (g'-r')_0 < 0.0$ and $0.8 < (u'-g')_0 < 1.5$.  
BHB candidates that fall outside the selection box in Fig.\
\ref{fig:photss} were objects originally selected by ``model'' magnitudes
from the Early Data Release; here we plot Data Release 1 Petrosian
magnitudes that we find are better behaved at bright magnitudes.  The
surface density of the SDSS-selected BHB candidates is 1 deg$^{-2}$.  
There is no overlap of these objects with the 2MASS-selected sample, even
though both samples cover the same region of sky.

        We looked up available 2MASS photometry for the SDSS-selected BHB
candidates and found matches for 188 of the 194 objects. A handful of SDSS
stars satisfy the 2MASS-selection in $J_0$ and $(J-H)_0$ but are rejected
by $(H-K)_0$.  Thus the lack of overlap between the 2MASS- and
SDSS-selected samples is likely due to the extreme uncertainties in 2MASS
colors for the fainter 16th and 17th magnitude SDSS stars (see Fig.\
\ref{fig:photss}).  Interestingly, errant G-type stars found in the SDSS
sample are cleanly identified by 2MASS photometry as bright and red
$(J-H)_0\simeq0.35$ stars. This comparison suggests that some bright (15th
to 16th magnitude) SDSS stars are likely saturated, and thus have
erroneous reported magnitudes. The on-line documentation for the SDSS data
archive now describes a series of flags that can be used to avoid such
saturated objects.

\begin{figure}
 \centerline{\includegraphics[width=2.5in]{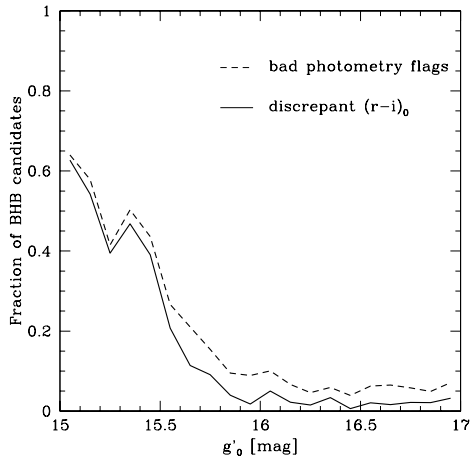}}
 \figcaption{ \label{fig:sat}
	Fraction of all A-colored stars in the SDSS DR2 equatorial region
with bad photometry flags ({\it dashed line}) or discrepant
$(r'-i')_0>0.3$ color ({\it solid line}).}
\end{figure}

	To understand our completeness for BHB stars requires a better
understanding of the SDSS saturation problem.  We start by selecting all
stars with A-type colors along the celestial equator in SDSS Data Release
2 \citep{abazajian04}.  We find that saturated objects have discrepant
$(r'-i')_0$ colors for A-type stars.  The solid line in Fig.\
\ref{fig:sat} shows the fraction of objects with discrepant
$(r'-i')_0>0.3$.  We then re-select all A-colored stars, but this time
using the photometry flags to select objects only with clean photometry.  
The dashed line in Fig.\ \ref{fig:sat} shows the fraction of objects with
clean photometry.

	Figure \ref{fig:sat} shows that half of all A-colored stars with
$15<g'_0<15.5$ have erroneous photometry and are not A stars at all.  For
these objects to have A-type colors in $(u'-g')_0$ and $(g'-r')_0$ but not
in $(r'-i')_0$ suggests that the $g'$ band is saturated.  This result also
suggests that half of the real A-colored stars may be missing in this
magnitude range.  Selecting for clean photometry removes the erroneous
objects, but may also reduce the completeness of the sample. The fraction
of discrepant A-colored stars drops to $\sim$10\% at $g'_0=15.75$ (see
Fig.\ \ref{fig:sat}), and the clean photometry selection maintains this
level of apparent incompleteness to $g'_0=17$.

\subsection{Spectroscopic Observations}

        During the fall 2003 observing season we obtained a spectrum for
each BHB candidate in the 2MASS- and SDSS-selected samples. Spectroscopic
observations were obtained with the FAST spectrograph \citep{fabricant98}
on the Whipple 1.5 m Tillinghast telescope.  We used a 600 line mm$^{-1}$ 
grating and a 2 arcsec slit to
obtain a resolution of 2.3 \AA\ and a spectral coverage from 3400 to 5400
\AA. Typical signal-to-noise ($S/N$) ratios were 30/1 in the continuum for
objects brighter than 16th magnitude, decreasing to $S/N$=15/1 for the
17th magnitude objects. This $S/N$ is adequate to measure the Balmer lines
and the Balmer jump -- the primary surface-gravity indicators we employ
for BHB stars. Paper I contains details of the data reduction.  We measure
spectral types and radial velocities, and derive metallicities, effective
temperatures, and surface gravities, from the spectra of the total sample
of 284 objects.

\subsection{BHB Classification}

	The major difficulty in using BHB stars as probes of Galactic
structure is the need to distinguish reliably between low surface-gravity
BHB stars and higher surface-gravity A-type dwarfs and blue stragglers.  
Although investigators once thought blue stragglers were a minor component
of the halo population, recent studies \citep{norris91,preston94,
wilhelm99b,brown03,clewley04} demonstrate that a surprisingly large
fraction of faint stars in the color range associated with BHB stars are
indeed high-gravity stars, many of which are blue stragglers
\citep{preston00,carney05}.

\begin{figure}
 \includegraphics[width=3.35in]{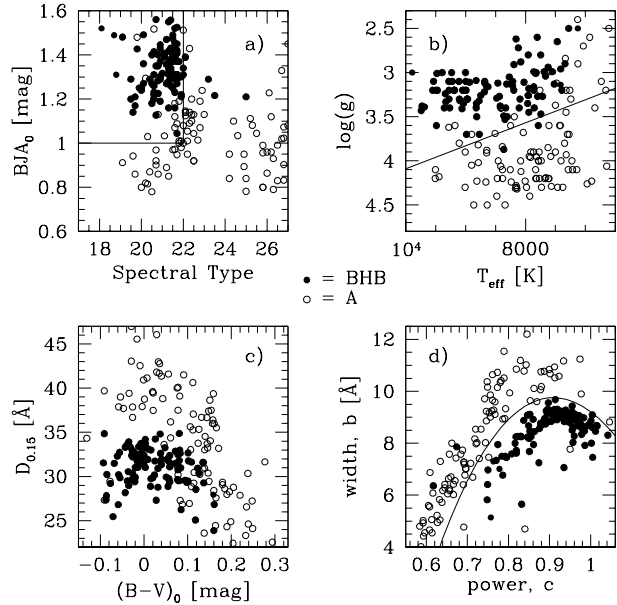}
 \figcaption{ \label{fig:bhball}
        The four BHB classification techniques applied to our sample:  (a)
the modified \citet{kinman94} method, (b) the \citet{wilhelm99a} method, (c)
the \citet{clewley02} $D_{0.15}$-Color method, and (d) the
\citet{clewley02} Scale width-Shape method.  Solid circles mark the BHB
stars; open circles mark the high surface-gravity A-type stars. }
\end{figure}

	Our classification of BHB stars is identical to the approach
described in Paper I.  In brief, we apply the techniques of
\citet{kinman94}, \citet{wilhelm99a}, and \citet{clewley02} to identify
low surface-gravity BHB stars.  We identify objects that satisfy three or
more of the four classification techniques as BHB stars (see Fig.\
\ref{fig:bhball}).  We find a total of 118 BHB stars across our 175
deg$^2$ survey region.

\subsection{Sample Selection Efficiencies}

\begin{deluxetable}{lccccc}		% SELECTION EFFICIENCIES
\tablewidth{3.25in}
\tablecaption{SELECTION EFFICIENCIES\label{tab:sel}}
\tablecolumns{6}
\tablehead{
	\colhead{Sample} & \colhead{$N_{stars}$} &
	\colhead{BHB} & \colhead{other A} &
	\colhead{B} & \colhead{F}
}
	\startdata
2MASS &	~90 & 38\% & 40\% & 7\% & 15\% \\
SDSS &	167 & 50\% & 42\% & 4\% & ~4\% \\
	\enddata 
 \end{deluxetable}

	Table \ref{tab:sel} summarizes sample selection efficiencies.
	The 2MASS-selected sample contains 34 BHB stars (out of 90
candidates) for a net selection efficiency of 38\%.  The total number of
A-type stars is about twice the number of BHB stars, or 78\% of the
2MASS-selected sample.  Of the remaining non A-type objects, 7\% of the
stars in the 2MASS-selected sample are B-type stars; 15\% of the stars in
this sample are F-type stars.

	The SDSS-selected sample contains 84 BHB stars (out of 167
candidates) for a net selection efficiency of 50\%.  We ignore the 27
G-type stars in this calculation as these stars can presumably be rejected
by saturation flags.  The total number of A-type stars is about twice the
number of BHB stars, or 92\% of the SDSS-selected sample (excluding 
the G-types).  Of the remaining non A-type objects, 4\% of the stars in 
the SDSS-selected sample are B-type stars;
4\% of the stars in this sample are F-type stars.

	\citet{sirko04a} have recently published a ``stringent'' color
selection for BHB stars.  Applying the stringent color cut to our full
SDSS-selected sample would yield 55 BHB stars selected from 81 candidates
for a net selection efficiency of 68\%, but a completeness of only 65\%
compared to the full SDSS-selected sample.

\subsection{Unusual Objects}

	In Paper I we identified a number of unusual objects, including
white dwarfs, subdwarfs, and B-type stars, within our survey of blue stars
in the halo.  The 2MASS- and SDSS-selected samples, by comparison, contain
a handful of B stars, but do not include any white dwarfs or subdwarfs.  
The lack of white dwarfs may be explained by the more restrictive color
selection we used for the 2MASS- and SDSS-selected samples.  Moreover, the
B-type stars in the 2MASS- and SDSS-selected samples are almost entirely
late B8 and B9 stars.  These late B-types are potentially all hot
horizontal-branch stars, but are very difficult to classify because the
horizontal branch crosses the main sequence at this location in the H-R
diagram.

        The earliest B-type star in our samples is \hbox{CHSS 1645},
classified as B7.  As the earliest B-type star in our samples, \hbox{CHSS
1645} is the most likely object to be a true B star rather than a hot
horizontal-branch star. Assuming \hbox{CHSS 1645} has solar metallicity,
with $M_V\sim-0.6$ \citep{allen00} and $(V-J)=-0.3$ \citep{kenyon95},
we estimate that it is located 6 kpc below the Galactic plane. This places
\hbox{CHSS 1645} among the class of stars known as ``run-away B-type''
stars. The star \hbox{CHSS 1645} is located at $b=-60^{\circ}$, hence its
+73 km s$^{-1}$ radial velocity points predominantly in the negative $z$
direction perpendicular to the plane of the Galaxy. If its radial
velocity is the majority of its full space motion, it takes 10$^8$ years
for \hbox{CHSS 1645} to travel 6 kpc from the Galactic plane. A B7 star
has $\sim$4 $M_{\odot}$ \citep{allen00} and a lifetime $\sim$2$\times10^8$
yr \citep{bowers84}. Thus \hbox{CHSS 1645}, a likely run-away B7 star, has
a lifetime consistent with its travel time from the disk.

\section{GLOBAL PROPERTIES}

	To map the Galactic halo requires knowing the intrinsic
luminosities of BHB stars.  BHB stars are standard candles with
luminosities that depend on effective temperature (color) as well as
metallicity.  We begin by discussing the physical basis of the BHB
luminosity dependence on color (Section 3.1).  We then present the
observed distribution of metallicities derived from our spectra (Section
3.2).  Using our colors and metallicities, we compute intrinsic
luminosities for our field BHB stars and investigate their spatial
distribution (Section 3.3).  Finally, we investigate the mean Galactic
rotation of our halo samples (Section 3.4).

\subsection{BHB Luminosity-Color Dependence}

	BHB stars share a common physical origin.  They are stars that
have evolved off the red giant branch and are burning helium in their
cores with a hydrogen burning shell. The bolometric luminosity of a BHB
star depends on the core mass, the stellar mass, and the metallicity
\citep[e.g.][]{demarque00}.  More massive BHB stars have larger
hydrogen-rich envelopes and are cooler than less massive BHB stars. The
variation of effective temperature with stellar mass yields a robust
relation between optical luminosity and B-V color: bluer BHB stars are
fainter than red BHB stars.

\begin{figure}
 \includegraphics[width=3.25in]{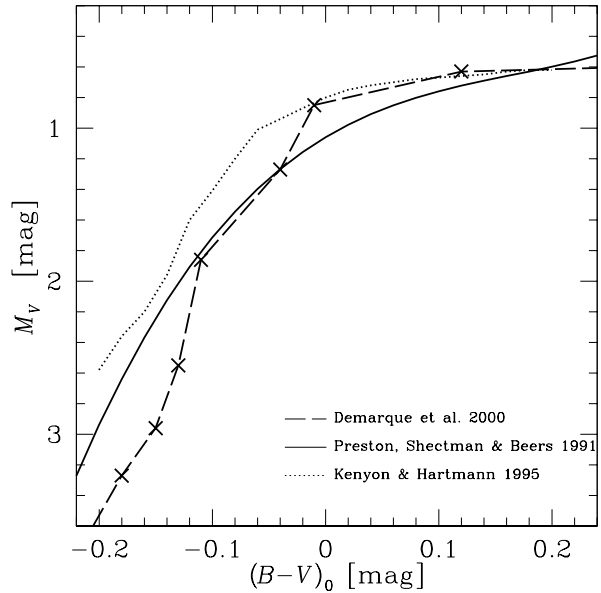}
 \figcaption{ \label{fig:models}
	BHB $M_V$-color relations.  The solid line shows the empirical
\citet{preston91} relation derived from globular clusters, the dashed line
shows the theoretical relation derived from the \citet{demarque00} stellar
models, and the dotted line shows our toy model of bolometric corrections
\citep{kenyon95} for a constant luminosity star.}
\end{figure}

	\citet{preston91}, hereafter PSB, provide an empirical BHB
luminosity-color relation from a fit to fifteen globular cluster BHBs.  
Fig.\ \ref{fig:models} shows the PSB $M_V$-color relation as a solid
line.  The relation is normalized to [Fe/H]=$-2.3$ using the
$M_V$-metallicity relation described below.  Globular clusters exhibit a
wide range of BHB morphologies, evident in the 0.25 mag scatter of the
points in PSB's Fig.\ 5.  Because the large scatter may result from the
physics of globular clusters rather than from the physics of BHB stars, we
next consider a theoretical $M_V$-color relation.  The theoretical
$M_V$-color relation provides a physical basis for the empirical PSB
relation.

	To construct a theoretical $M_V$-color relation, we use the
horizontal branch evolutionary tracks from \citet{demarque00}.  For models
with $Z=10^{-4}$ (equivalent to [Fe/H]$=-2.3$) we adopt the luminosity at
time zero and derive colors and bolometric temperatures from published
tables \citep{kenyon95,green96,lejeune98}. The resulting $M_V$-color
relation for $Z=10^{-4}$ is the dashed line in Fig.\ \ref{fig:models}.  
The theoretical $M_V$-color relation is remarkably similar in shape to the
empirical PSB relation in the (\bv)$_0>-0.1$ region covered by our BHB
star samples.

	As a consistency check, we plot a third line in Fig.\
\ref{fig:models} that is simply the bolometric correction for a star with
constant luminosity.  We use the bolometric corrections for main sequence
stars from \citet{kenyon95}, and add 0.5 mag to match the bolometric
correction to the empirical and theoretical $M_V$-color relations at the
red end.  Interestingly, the shape of the \citet{kenyon95} bolometric
corrections is similar to both the empirical and theoretical BHB
$M_V$-color relations, except that the slope of the bolometric correction
curve is too shallow at the blue end.  We expect this systematic
difference because blue BHB stars are intrinsically less luminous
than red BHB stars; we have assumed a constant luminosity star.  
Bolometric corrections from \citet{green96} and \citet{lejeune98} yield
similar results, with a typical scatter of 0.1 - 0.2 mag.  This toy model
shows that the primary ingredient in the BHB $M_V$-color relation is the
bolometric correction for BHB stars.

%a difference in zero point
%simply shifts our BHB luminosities and distances.  Because our analyses
%rely on relative distances, we are concerned primarily with the shape of
%the $M_V$-color relation.

	Thus the physics common to all BHB stars leads to a general BHB
$M_V$-color relation, albeit with an intrinsic spread resulting from age
and metallicity.  The $M_V$-color relation depends on age because the
luminosity and effective temperature of a BHB star evolve with time. The
$M_V$-color relation has a well known dependence on metallicity, but
\citet{demarque00} argue for an additional spread in $M_V$ at a given
metallicity due to BHB morphology.  The morphology effect is strongest for
a metal poor [Fe/H]$<-2$ BHB with blue morphology (HB type index=+1).  
According to \citet{demarque00}, a metal poor, blue BHB is actually
$\sim$0.1 mag brighter than the standard luminosity-metallicity relation
predicts.  We conclude that the BHB $M_V$-color relation has an {\it
intrinsic shape} due to the physics of the horizontal branch, with an
intrinsic spread of 0.1 - 0.2 mag.  For purposes of discussion, we use the 
empirical PSB $M_V$-color relation to
estimate BHB luminosities in the rest of this section.

\subsection{Metallicities}

	We measure metallicities for BHB stars as described in Paper I.  
We use three different techniques:  the line indices of \citet{beers99},
the equivalent width of Ca {\sc ii} K plus a chi-square comparison between
metallic-line regions in synthetic and observed spectra
\citep{wilhelm99a}, and an optimization method that fits the entire
spectrum \citep{allende03}.  The three techniques are in good agreement
with 0.25 dex uncertainty \citep{brown03}.  The final metallicity is the
average of the three techniques; we adopt $\pm$0.25 dex as the error in
the final metallicity.

\begin{figure}
 \includegraphics[width=3.25in]{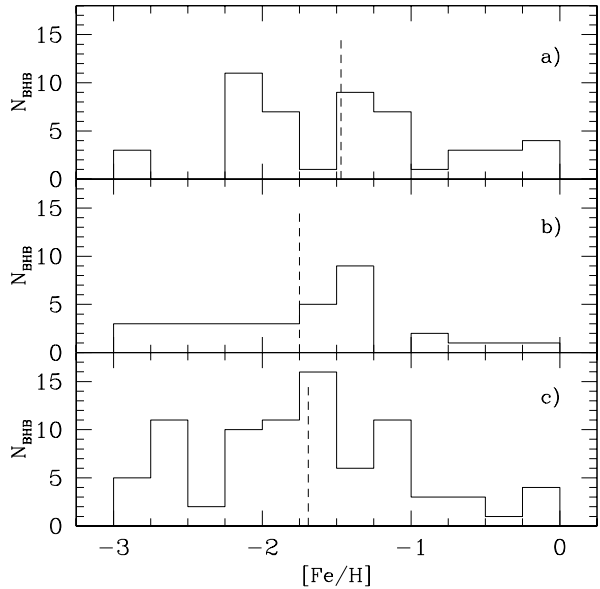}
 \figcaption{ \label{fig:fehbhb}
	Distribution of BHB [Fe/H] for (a) the Century Survey sample, (b)  
the 2MASS-selected sample, and (c) the SDSS-selected sample.  Dashed lines
indicate the median [Fe/H].  The error in [Fe/H] is 0.25 dex, the same 
size as our bins.}
\end{figure}

	Figure \ref{fig:fehbhb} plots the observed distribution of
metallicities we measure in the original Century Survey sample, the
2MASS-selected sample, and the SDSS-selected sample.  The median [Fe/H] of
the BHB samples are indicated by the dashed lines in Fig.\
\ref{fig:fehbhb} and range from [Fe/H]=$-1.47$ to $-1.75$.  Our
metallicity-measuring techniques are limited to the range $-3<$[Fe/H]$<0$,
and so peaks in Fig.\ \ref{fig:fehbhb} at [Fe/H]=$-3$ and 0 are likely
stars with lower/higher metallicities piling up at the limits.  We caution
that the distributions in Fig.\ \ref{fig:fehbhb} are the observed and not
the intrinsic distributions:  because BHB luminosities have a mild
dependence on metallicity, stars of different [Fe/H] are sampled from
different volumes of space (see below).  However, it is clear that our BHB
samples are predominantly metal-poor and therefore consistent with a halo
population.

	A Kolmogorov-Smirnov (K-S) two-sample test provides a simple way
to evaluate whether the different BHB samples are drawn from a common
parent distribution of [Fe/H].  The K-S two-sample test works by sorting
an observed quantity, such as metallicity, and then comparing the 
cumulative distributions of two different samples with one
another.  The likelihood is calculated for the null hypothesis that the
two distributions are drawn from the same parent distribution.  We test
the metallicity distributions in Fig.\ \ref{fig:fehbhb} in a pairwise
fashion, and find likelihood values ranging from 14\% to 60\%.  Thus our
BHB samples are consistent with the null hypothesis that the metallicity
distributions come from the same (halo) population.

\subsection{Spatial Distribution}

	We calculate luminosities for our field BHB stars using the
$M_V(BHB)$ relation from \citet{clewley04}.  This relation assumes the
{\it Hipparcos}-derived zero point, $M_V(RR) =0.77\pm0.13$ at [Fe/H] =
$-1.60$ \citep{gould98}, a $M_V$-metallicity slope $0.214\pm0.047$ based
on RR Lyrae stars in the Large Magellanic Cloud \citep{clementini03}, and
the PSB cubic relation in (\bv)$_0$ to provide the temperature correction.  
Although the PSB $M_V$-color relation was derived for globular cluster BHB
stars, the {\it shape} of the relation reflects the physics common to all
BHB stars, as explained above.  Note that we do not measure (\bv)$_0$
directly.  For the SDSS sample, we are able to make accurate estimates of
(\bv)$_0$ from SDSS colors.  For the 2MASS sample, we use 2MASS photometry
and Balmer line strengths to estimate (\bv)$_0$ as described in Paper I.  
We refer to these (\bv)$_0$ estimates as $BV0$.  From the derived
luminosities we compute distances.  We expect the relative distances of
our BHB stars have a precision of $\sim$6\%.

\begin{figure}
 \includegraphics[width=3.35in]{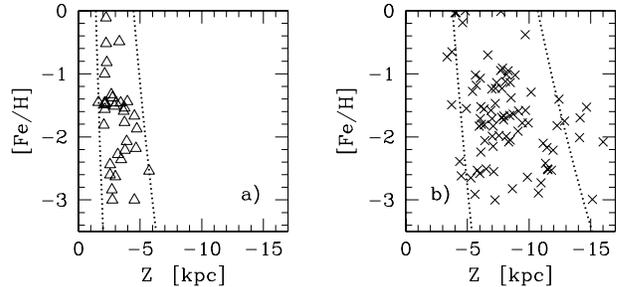}
 \figcaption{ \label{fig:limfeh}
	Distribution of [Fe/H] versus $z$ for (a) the 2MASS- and 
(b) the SDSS-selected samples.  The dotted lines represent the magnitude limits 
for a BHB star at the median Galactic latitude $b=-57^{\circ}$ and at the 
median color $BV0=0.03$ of the samples.  Some stars fall beyond the 
``limits'' because there is a spread of Galactic latitude and color in our 
samples.}
\end{figure}

	Because the luminosity of a BHB star is dependent on metallicity,
the depths reached by our flux-limited samples are dependent on
metallicity.  Fig.\ \ref{fig:limfeh} shows the distribution of [Fe/H] for
the 2MASS- and SDSS-selected samples as a function of $z$, the distance
above or below the Galactic plane.  The dotted lines represent the
magnitude limits for a BHB star at the median Galactic latitude
$b=-57^{\circ}$ and at the median color $BV0=0.03$ of our samples.  
Metal-rich BHB stars are intrinsically fainter than metal-poor BHB stars.  
Fig.\ \ref{fig:limfeh} shows that we sample BHB stars with [Fe/H]$=-1$ to
82\% the depth of BHB stars with [Fe/H]$=-3$.

\begin{figure}
 \includegraphics[width=3.35in]{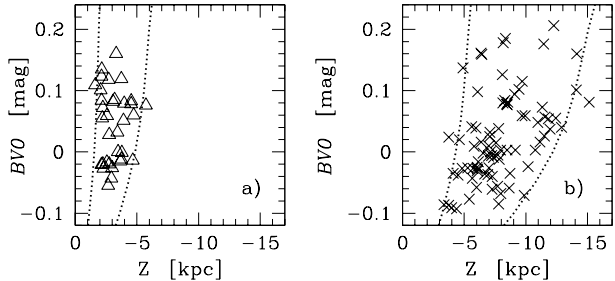}
 \figcaption{ \label{fig:limbv0}
	Distribution of $BV0$ versus $z$ for (a) the 2MASS- and (b) the
SDSS-selected samples.  The dotted lines represent the magnitude limits
for a BHB star at the median Galactic latitude $b=-57^{\circ}$ and at the
median metallicity [Fe/H]$=-1.7$ of the samples.  Some stars fall beyond
the ``limits'' because there is a spread of Galactic latitude and
metallicity in our samples.}
\end{figure}

	The BHB luminosity dependence on color is stronger than the
dependence on metallicity.  Thus there is a strong selection bias with
stellar color.  In Fig.\ \ref{fig:limbv0} we plot the distribution of
$BV0$ color for the 2MASS- and SDSS-selected samples as a function of $z$.
The dotted lines represent the magnitude limits for a BHB star at the 
median
Galactic latitude $b=-57^{\circ}$ and at the median metallicity
[Fe/H]$=-1.7$ of our samples.  Fig.\ \ref{fig:limbv0} shows that we sample
BHB stars with $BV0=-0.1$ to only 64\% of the depth that we detect BHB
stars with $BV0=+0.1$.  The intrinsically faintest BHB stars are the
bluest BHB stars hooking down off the horizontal branch in an H-R diagram.  
These faint BHB stars are sampled in a smaller volume than the more
luminous BHB stars in our samples.

\begin{figure}
 \includegraphics[width=3.25in]{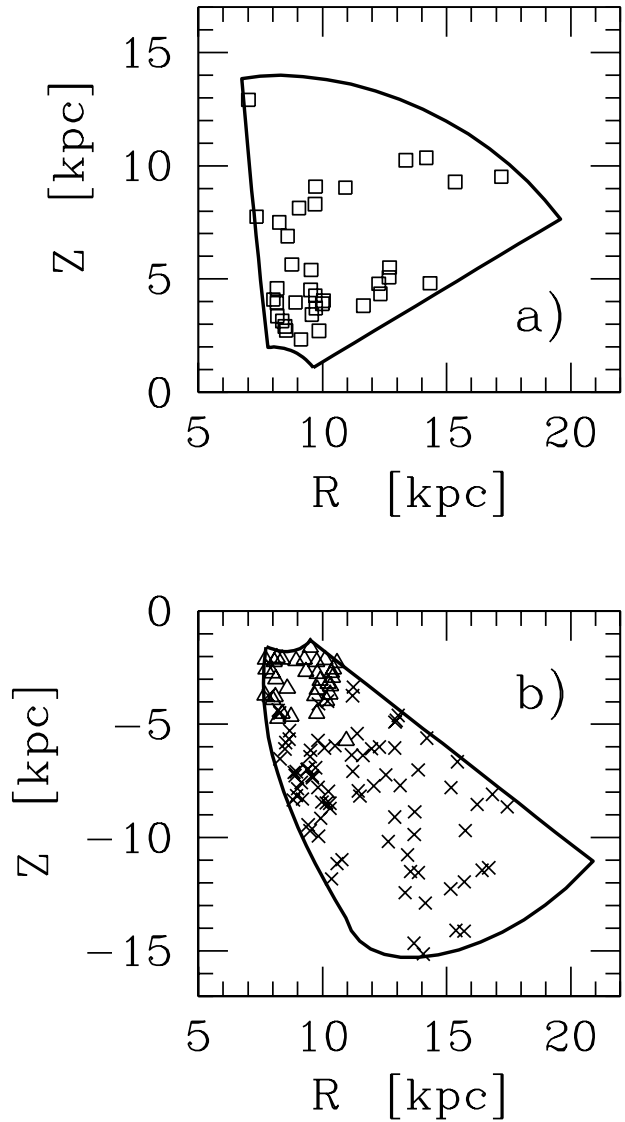}
 \figcaption{ \label{fig:rz}
	Distribution of BHB stars in distance from the Galactic center
along the Galactic plane, $R$, and distance above/below the Galactic
plane, $z$.  Panel (a) shows the BHB stars in the original Century Survey
slice.  Panel (b) shows the BHB stars in the 2MASS-selected (triangles)
and SDSS-selected (crosses)  samples.  The solid lines indicate
heliocentric distance limits of 2 kpc and (a) 14 kpc or (b) 17 kpc.}
\end{figure}

	We now plot the spatial distribution of the original Century
Survey, the 2MASS-selected, and the SDSS-selected BHB samples (Fig.\
\ref{fig:rz}).  Spatial distribution is traditionally displayed in a wedge
plot for survey slices like ours.  However, a wedge plot is inappropriate
in the context of the Galaxy, where a slice in celestial coordinates cuts
across varying Galactic latitudes.  The density of halo and disk
populations is a strong function of both $R$, the distance along the
Galactic plane, and $z$ \citep[e.g.][]{siegel02}.  Thus in Fig.\
\ref{fig:rz} we plot the distribution of BHB stars as a function of $R$
and $z$.  Indeed, the observed distribution of BHB stars clearly depends
on both $R$ and $z$.

	The 2MASS- and SDSS-selected BHB candidates sample complementary
ranges of distances, but the overlap is unfortunately minimal.  The 2MASS
and SDSS catalogs formally overlap between 15 and 15.5 magnitude.  
However, in this magnitude range, the 2MASS BHB selection suffers from
incompleteness due to large color errors and the SDSS BHB selection likely
suffers from incompleteness due to saturation problems.

\subsection{Mean Galactic Rotation}

        There is wide disagreement in the literature on whether the
stellar halo rotates significantly. Previous surveys have found evidence
for (1) no halo rotation \citep{layden96, gould98, martin98, gilmore02,
sirko04b}, (2) a small prograde rotation \citep{chiba00}, and (3)
retrograde rotation \citep{majewski92,majewski96,spagna03,kinman05}.  
Interestingly, all the measurements of retrograde rotation come from
surveys of the north Galactic pole.  By comparison, the measurements of no
rotation come from surveys covering many directions in the sky.

	Our BHB samples cover a wide range of Galactic latitude and
longitude, and so provide us with a ``fair'' sample of the halo.  
Although high Galactic latitude stars are not ideal for measuring the
rotation of the stellar halo, the 2MASS- and SDSS-selected samples include
a number of stars near $l\sim90^{\circ}$ that are sensitive to a
systematic rotation of the halo.

\begin{figure}
 \includegraphics[width=3.35in]{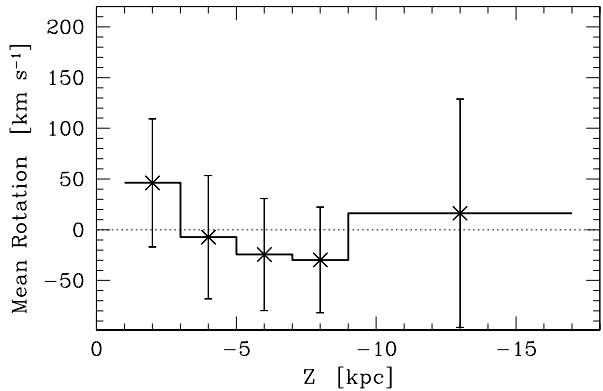}
 \figcaption{ \label{fig:rotz}
	Galactic rotation velocity of the 2MASS- and SDSS-selected BHB 
stars, binned by distance below the Galactic plane.}
\end{figure}

	Figure \ref{fig:rotz} shows the mean rotation velocity of the
2MASS- and SDSS-selected BHB stars as a function of $z$.  Each bin
includes $\sim$25 BHB stars that are first corrected to the local standard
of rest \citep{dehnen98} and then evaluated by the methodology of
\citet{frenk80}.  We assume a solar rotation of 220 km s$^{-1}$. We expect
some contamination from the thick disk in the nearest BHB stars, and
indeed there is a hint of prograde rotation in the $1<|z|<3$ kpc bin.  
The mean rotation velocities of the $3<|z|<15$ kpc stars are, however,
consistent with no rotation.  The rotation velocity of the combined sample
of 2MASS- and SDSS-selected BHB stars is $-4\pm30$ km s$^{-1}$.  The
velocity dispersion of the BHB stars is 108 km s$^{-1}$, also consistent
with a halo population.

	The BHB stars from the original Century Survey sample cover a
similar range of $z$.  However, the Century Survey BHB stars are located
towards the Galactic anti-center $l\sim200^{\circ}$ and towards the north
Galactic pole $b\gtrsim60^{\circ}$.  Thus the Century Survey stars provide
very little leverage on halo rotation.  When we include the Century Survey
stars in the mean rotation velocity calculation, we find they add $\sim10$
km s$^{-1}$ of retrograde rotation to the bins.  We conclude the mean
rotation velocities remain fully consistent with no halo rotation within
their errors.

\section{THE LUMINOSITY FUNCTION OF BHB STARS}

	Knowledge of the intrinsic distribution of luminosities of field
BHB stars is important for interpreting maps of the Galactic halo.  
Knowledge of the luminosity function is also important for understanding
the intrinsic properties of field BHB stars that cover a broad range of
observed magnitude, color, and metallicity.  The luminosity function
describes the number of stars per unit volume in the luminosity interval
$L$ to $L+dL$.  We describe the method we use to calculate the BHB
luminosity function (Section 4.1), and discuss the role of the $M_V$-color
relation in our result (Section 4.2).  We compare the luminosity function
we determine for our field BHB stars (Section 4.3) with luminosity
functions derived from globular clusters with BHBs (Section 4.4).

\subsection{Calculating the Luminosity Function}

	We calculate the luminosity function of our field BHB stars using
the non-parametric maximum-likelihood method of \citet{eep}.  The
\citet{eep} maximum-likelihood method is commonly used to calculate the
luminosity function of galaxies in galaxy redshift surveys.  We now apply
this method to our survey of BHB stars in the Galactic halo.  The method
does not simply count the numbers of stars at different luminosities, but
weights the contribution of each star by the relative volume in which it
can be observed in a flux-limited sample.  Specifically, the probability
of a star at distance $d$ falling into the luminosity range $[L,L+dL]$ is
equal to the luminosity function at $L$ divided by the number density of
stars one expects to see in a flux-limited survey at distance $d$.  The
maximum-likelihood method works by maximizing the sum of these
probabilities and solving for the best-fitting luminosity function.

	The density terms drop out in the maximum-likelihood formalism
with two notable consequences.  First, the maximum-likelihood method is
unbiased by systematic density variations.  The maximum-likelihood method
does not require knowledge of the halo density distribution $\rho(R,z)$;
it only requires that the luminosity function is independent of position
in the sampled volume.  Secondly, the absolute normalization of the
luminosity function is lost and requires a separate computation.  Because
stellar density varies with position in the Milky Way and because our
samples are too sparse to fit the density profile directly, we compute
only the {\it form} of the luminosity function, and arbitrarily normalize
the luminosity functions to unity.

\subsection{The Role of the $M_V$-color Relation}

%	At first glance, it may appear that the maximum-likelihood method
%will simply reproduce the relations we use to derive the BHB luminosities
%in the first place.  This is incorrect because we have no idea how the
%$M_V$ relations are {\it populated}. 

\begin{figure}
 \includegraphics[width=3.25in]{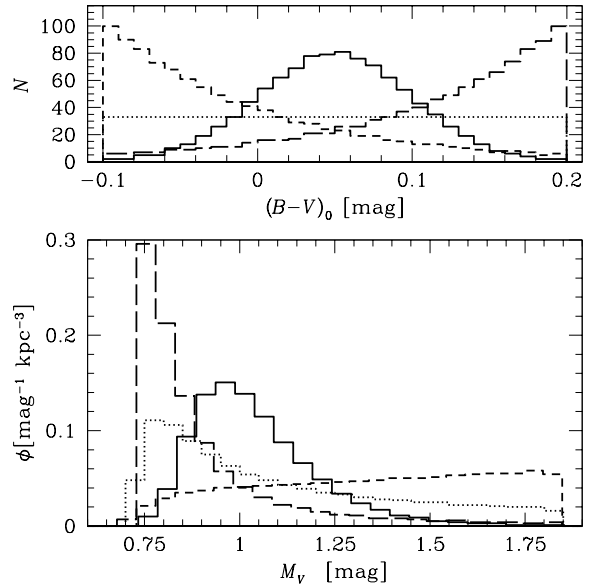}
 \figcaption{ \label{fig:simlf}
	We recover the intrinsic luminosity functions (lower panel) for
four simulated intrinsic color distributions (upper panel), in this case
using the PSB $M_V$-color relation.  The distribution of colors is the
crucial element for the luminosity function.}
\end{figure}

	The $M_V$-color relations (Fig.\ \ref{fig:models}) specify only
how a particular color maps to a particular $M_V$.  The distribution of
colors is not at all specified by the $M_V$-color relation.  The
distribution of colors is the crucial element of the luminosity function.
To illustrate this point, Fig.\ \ref{fig:simlf} plots luminosity functions
calculated for the following four {\it intrinsic} color distributions:  
(1) an uniform color distribution, (2) a Gaussian color distribution
centered at (\bv)$_0$=0.05 with $\sigma=0.05$ mag, and (3) exponential
color distributions with scale length 0.1 mag peaking in the red and (4)
the blue.  Each model color distribution contains 1000 objects.  For
purposes of this calculation, we derive the intrinsic BHB luminosities
using the PSB $M_V$-color relation, though our results are nearly
identical for the other $M_V$-color relations in Fig.\ \ref{fig:models}.  
The bottom panel of Fig.\ \ref{fig:simlf} shows the luminosity functions
resulting from the four color distributions.

	It is clear from Fig.\ \ref{fig:simlf} that the BHB luminosity
function depends dramatically on the distribution of BHB colors.  Each
simulated luminosity function in Fig.\ \ref{fig:simlf} has a different
shape, some with narrow distributions, others with long tails extending to
faint luminosities.  Moreover, the characteristic peaks of the luminosity 
functions vary in luminosity and total number of stars.

	Even though colors are the primary indicator of BHB luminosity, we
cannot compare raw distributions of colors because our field BHB stars
have different luminosities and thus sample different volumes of space.  
To derive intrinsic properties requires knowing the luminosity function of
our field BHB stars.

\subsection{The Field BHB Luminosity Function}

\begin{figure}
 \includegraphics[width=3.25in]{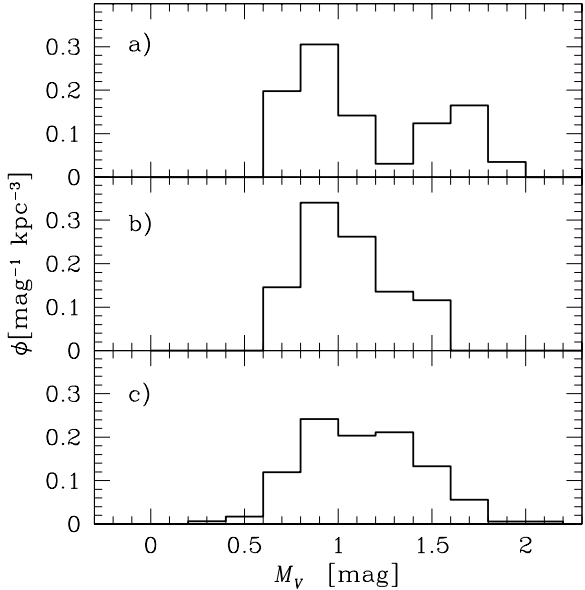}
 \figcaption{ \label{fig:lf}
	Luminosity functions of halo BHB stars in (a) the Century Survey 
sample, (b) the 2MASS-selected sample, and (c) the SDSS-selected sample.  
The normalization is scaled so that the areas under the curves are equal 
to one.}
\end{figure}

\begin{figure}
 \includegraphics[width=3.35in]{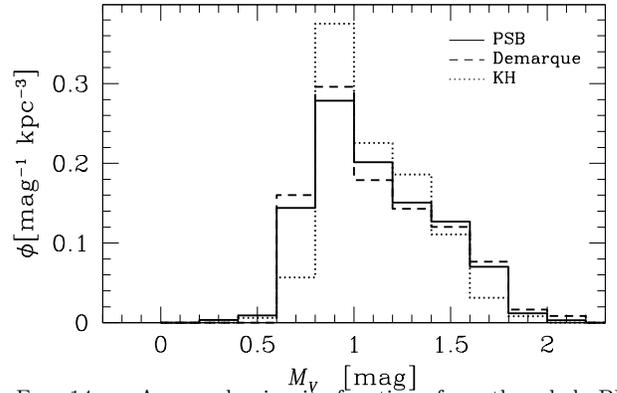}
 \figcaption{ \label{fig:av3}
	Average luminosity function of our three halo BHB star samples,
calculated using (1) the empirical PSB $M_V$-color relation (solid line),
(2) the theoretical \citet{demarque00} BHB models (dashed line), and (3)  
our toy model of bolometric corrections \citep{kenyon95} for a constant
luminosity BHB star.}
\end{figure}

\begin{figure}
 \includegraphics[width=3.25in]{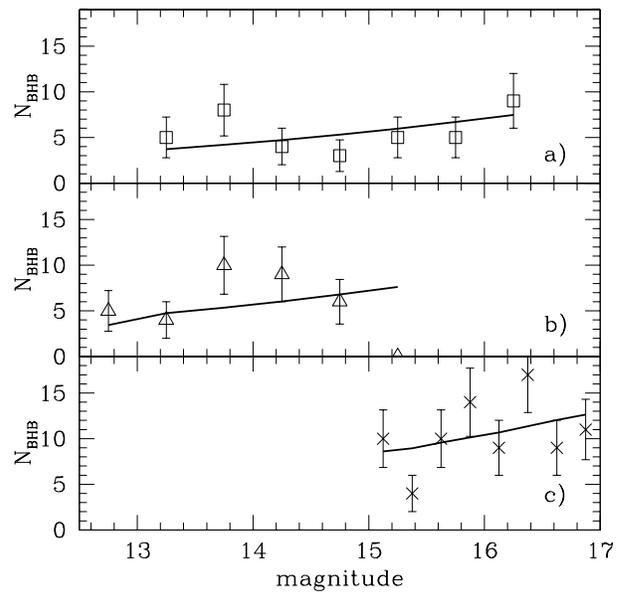}
 \figcaption{ \label{fig:maglf}
	Apparent magnitude distribution of BHB stars in (a) the Century
Survey sample, (b) the 2MASS-selected sample, and (c) the SDSS-selected
sample.  Extinction-corrected magnitudes are (a) $V_0$, (b) $J_0$, and
(c) $g'_0$.  Solid lines indicate the number of BHB stars predicted by our 
derived luminosity functions assuming a $r^{-2.5}$ halo density profile,
normalized to the number of stars in each sample.}
\end{figure}

	Figure \ref{fig:lf} shows the luminosity function of the original
Century Survey, the 2MASS-selected, and the SDSS-selected samples,
determined from the observations of color, metallicity, and apparent
magnitude.  We use 0.2 magnitude wide bins to encompass any uncertainties
in the $M_V$ derivation.  Luminosities are derived with the previously
stated \citet{clewley04} relation that uses the PSB $M_V$-color relation.
All three luminosity functions display the same general shape:  a steep
rise at bright luminosities, a peak between $0.8 < M_V < 1.0$, and a tail
at faint luminosities.

	We perform K-S tests to measure the likelihood that our field BHB
luminosity functions are drawn from the same parent population.  The K-S
test applies to unbinned distributions; we thus multiply the luminosity
functions (Fig.\ \ref{fig:lf}) by the number of objects in the samples and
distribute the $M_V$'s uniformly across each bin.  The resulting
likelihoods range from 37\% to 57\%, suggesting that our BHB samples share
a common parent population.  These likelihoods also mean that the bimodal 
distribution of luminosities in the Century Survey sample
(Fig.\ \ref{fig:lf}a) is not statistically significant.  Interestingly,
all three samples share the same median $M_V\simeq1.0$ mag.  This
agreement is rather remarkable, given the independent photometry of the
three samples, and suggests that the Century Survey, 2MASS, and SDSS have
consistent photometry.

	Because the K-S tests suggest our three BHB samples are drawn from
the same halo population, we average the three BHB samples to obtain a
more robust measure of the field BHB luminosity function.  We multiply
each luminosity function in Fig.\ \ref{fig:lf} by the total number of BHB
stars in each sample, sum the luminosity functions, and then divide the
result by the grand total of BHB stars.  The result is plotted as the
solid histogram in Fig.\ \ref{fig:av3}.

	We re-compute the luminosity functions using the $M_V$-color
relations derived from (1) \citet{demarque00} and (2) \citet{kenyon95}. We
show the results as (1) dashed and (2) dotted histograms in Fig.\
\ref{fig:av3}.  Because we are interested in the shape of the luminosity
function, we adjust the zero-points of the $M_V$-color relations derived
from \citet{demarque00} and \citet{kenyon95} to match the zero-point of
the PSB relation.  Interestingly, the theoretical \citet{demarque00} and
empirical PSB curves have very similar shape: a K-S test gives a 99\%
likelihood for the two samples to share a common distribution.  Thus the
shapes of the theoretical and empirical $M_V$-color relations are similar
enough to have no apparent effect on the shape of the final BHB luminosity
function.

	Knowledge of the BHB luminosity function allows us, in theory, to
solve for the BHB density distribution.  In practice, our relatively
sparse samples do not provide an adequate constraint (see Fig.\
\ref{fig:maglf}).  We note that the halo power laws and scale lengths
published in \citet{siegel02} yield reduced $\chi^2\sim1$ and so appear
consistent with the distribution of our field BHB stars.

	To check of the veracity of our average BHB luminosity function,
we use the luminosity function in Fig.\ \ref{fig:av3} (solid histogram) to
calculate the expected apparent magnitude distributions of our BHB
samples.  Fig.\ \ref{fig:maglf} plots the observed number of BHB stars in
the original Century Survey, the 2MASS-selected, and the SDSS-selected
samples versus extinction-corrected apparent magnitude.  The errorbars
indicate $\sqrt{N}$ uncertainties.
	The solid lines in Fig.\ \ref{fig:maglf} are the number of BHB
stars predicted by the BHB luminosity function, assuming a $r^{-2.5}$
power-law density profile \citep{siegel02}.  To set the normalization, we
scale the predictions to the observed number of stars in each sample.  
There is good agreement in the predicted shape of the magnitude
distribution and the observations.  The one exception is the final 2MASS
bin with $15<J_0<15.5$.  We attribute the observed under-density to larger
photometric errors at faint magnitudes (see Fig.\ \ref{fig:photss}) that
scatter BHB stars out of the narrow 2MASS color-selection box, thereby
reducing our completeness.

% It is also possible that large scale 
% structure in the halo may account for some of the under-density.

\subsection{Comparison with Globular Clusters}

	An additional insight into our field BHB luminosity function is
provided by comparison with globular cluster data.  The purpose of this
comparison is not to suggest that the halo is made of disrupted globular
clusters.  Rather, because all BHB stars share a common physical basis, we
inquire whether they exhibit a common parent distribution of luminosities.  
Globular cluster BHB morphologies are known to vary widely because of
differences in metallicity, main sequence turn-off mass, and
``second-parameter'' effects.  We expect that our wide area surveys of the
halo will sample BHB stars from the full range of BHB morphologies.  We
now test whether field and globular cluster BHB stars share a similar or
different distribution of luminosities by comparing the shapes and median
$M_V$'s of the BHB luminosity functions.

\begin{figure*}
 \includegraphics[width=7.0in]{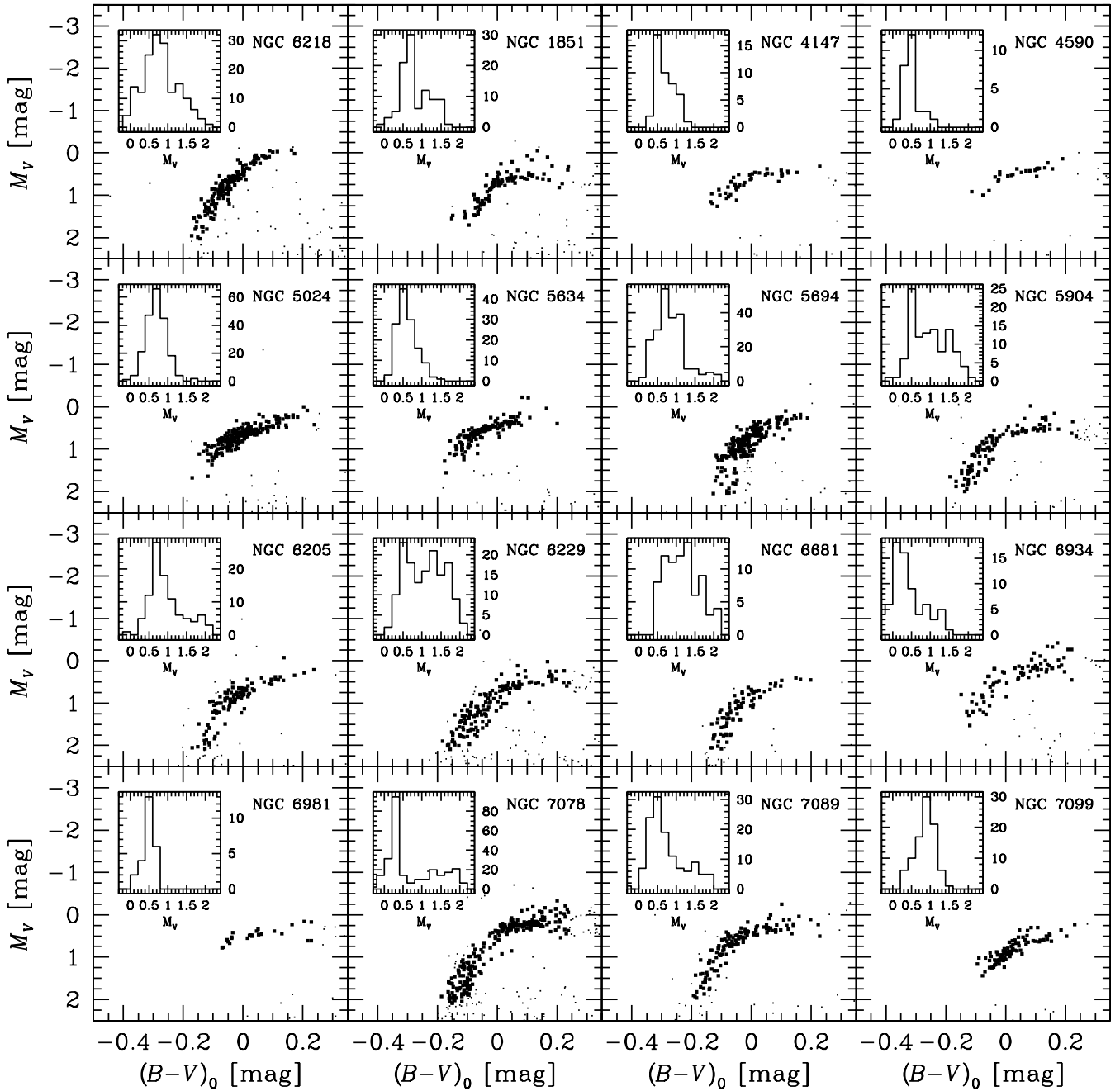}
 \caption{ \label{fig:gc}
	The extinction-corrected color-magnitude diagrams for M12
\citep[NGC 6218;][]{hargis04} and 15 additional globular clusters
\citep{piotto02}.  BHB stars are marked as bold points.  The luminosity
function of the BHB stars are shown in the inset panels.}
\end{figure*}

	In a brief example of the virtual observatory in action, we used
the NASA ADS system \citep{kurtz00} to locate suitable globular cluster
data for comparison with our field BHB samples.  We used the query
``globular cluster color magnitude diagram'' and required that there be
on-line data associated with the paper.  The first (most recent) paper
which met all our requirements is the study by \citet{hargis04} of M12
(NGC 6218).  We followed the data link to the CDS/VIZIER system
\citep{ochsenbein00}, from which the photometry data table was easily
downloaded.  Fig.\ \ref{fig:gc} shows the extinction-corrected
color-magnitude diagram for M12 (top left).  Stars on the BHB, marked as
bold points in Fig.\ \ref{fig:gc}, were selected by eye.  M12 is
relatively metal-poor [Fe/H] $\sim-1.4$, similar to our halo star samples
with median [Fe/H]$=-1.7$.

	The inset in Fig.\ \ref{fig:gc} shows the luminosity function of
BHB stars in M12.  There is considerable uncertainty in the distance
modulus and the metallicity of M12, with values ranging from
$(m-M)=14.22\pm0.11$ for [Fe/H]$=-1.14$ to $(m-M)=13.96\pm0.11$ for
[Fe/H]$=-1.61$.  We calculate absolute magnitudes using the distance
modulus $(m-M)=14.05\pm0.12$ appropriate for [Fe/H]$=-1.4$
\citep{hargis04} and extinction $E(B-V)=0.19\pm0.02$.  The median
extinction to our halo BHB stars, by comparison, is $E(B-V)=0.03$ mag. The
M12 BHB luminosity function has median $M_V=0.75$, a quarter of a
magnitude brighter than our field BHB samples.  The disagreement in median
$M_V$ is significant only at the $1\sigma$ level, however, since the M12
distance modulus, the M12 extinction correction $A_V$, and the $M_V(BHB)$
zero-point are all uncertain to $\pm0.1$ mag.

	The shape of the M12 BHB luminosity function is nearly identical
to the shape of our field BHB luminosity functions.  We use a K-S test as
before and calculate the likelihood that the M12 and our field BHB
luminosity functions are drawn from the same distribution.  Because our
goal is to compare the shapes of the luminosity functions, we match the
median $M_V$ of the observed and M12 samples when performing the K-S test.  
We find likelihoods ranging from 24\% to 44\% for our field BHB samples,
indicating that they likely share the same distribution of BHB
luminosities as the M12 BHB sample.

	\citet{piotto02} provide on-line data for 74 additional globular
clusters that we use for further comparison.  The data come from {\it
Hubble Space Telescope} $F439W$ and $F555W$ imaging from which
\citet{piotto02} derive de-reddened $B$ and $V$ magnitudes.  Because {\it
Hubble Space Telescope} targeted the centers of the globular clusters, not
all of the color-magnitude diagrams are as well sampled as the
\citet{hargis04} M12 data.  One third of the \citet{piotto02} globular
clusters have horizontal branches that are non-existent or too
sparsely populated to provide a meaningful comparison with our BHB
samples.  Of the remaining 51 globular clusters, we select the 15 with
$E(B-V)<0.1$ to minimize uncertainties in extinction.  These 15 globular
clusters span the range of metallicity $-2.3 <$ [Fe/H] $<-1.2$.

	Figure \ref{fig:gc} shows the extinction-corrected color-magnitude
diagrams for the 15 globular clusters from \citet{piotto02}.  We calculate
absolute magnitudes using the distance moduli and extinction values given
by \citet{piotto02}.  Interestingly, each of the 15 globular clusters has
a different median BHB $M_V$ than our BHB samples.  There is no
correlation with globular cluster metallicity.  The average globular
cluster median BHB luminosity is $M_V=0.7\pm0.25$.  Although a 0.3 mag
difference from our BHB samples is not formally significant, we expected
better agreement when averaging over this set of globular clusters.  In
Paper I, we noticed a similar $\sim$0.3 magnitude discrepancy between our
absolute magnitudes and the theoretical calculations for the zero age
horizontal branch.  For example, the \citet{demarque00} model discussed
here (Fig.\ \ref{fig:models}) is 0.1 mag brighter than the PSB relation.  
If the zero-point of our $M_V(BHB)$ relation is in error by 0.3 mag, then
our BHB stars are 14\% more distant than our current estimates.  Given the
strong dependence of BHB luminosity on color, it may be worth re-visiting
the PSB analysis and their zero point.

        The {\it shape} of the \citet{piotto02} globular cluster BHB
luminosity functions are in good agreement with our field BHB luminosity
functions. The globular cluster BHB stars are marked as bold points in
Fig.\ \ref{fig:gc}.  We select BHB stars by color/magnitude cuts, imposing
the same limits $(B-V)<0.24$ and $M_V<2.1$ as for our samples.  The insets
in Fig.\ \ref{fig:gc} show the luminosity functions of globular cluster
BHB stars.  We perform K-S tests on the shapes of the luminosity functions
as before, first matching the median $M_V$ of the globular cluster BHB
stars to our BHB samples.  The likelihoods that the BHB stars are drawn
from the same distribution range from 20\% to 80\%, with the exception of
two globular clusters.  NGC 6229 and NGC 7078 (M15) have significant
extended horizontal branches and thus a much broader distribution of BHB
luminosities than our field BHB samples. The K-S test yields a 10\%
likelihood for NGC 6229, which would only allow a very marginal rejection
of the null hypothesis of a common parent distribution.  A 1\% likelihood
is obtained for NGC 7078, indicating that its luminosity distribution is
not consistent with our field BHB samples.  We note that the metallicities
of NGC 6229 and NGC 7078 are [Fe/H]=$-1.43$ and $-2.25$, respectively.

	Of the sixteen globular clusters displayed in Fig.\ \ref{fig:gc},
fourteen (88\%) have luminosity functions consistent in shape with our
field BHB luminosity functions. Thus, under the assumption that the
$M_V$-color relation is intrinsic to stars on the BHB, we find that field
BHB stars in the halo and BHB stars in globular clusters with BHBs appear
to share a common distribution of luminosities.  The exception to this
conclusion are globular clusters with significant extended BHBs; we do not
see large numbers of extended BHB stars with (\bv)$_0<-0.1$ in our
samples.
	In the future, it would be useful to compare our field BHB
luminosity function with dwarf spheroidals and open clusters.

%	As a final test, we use the on-line data of \citet{lee03} to
%compare our BHB luminosity function with that of the Sextans dwarf galaxy.  
%We calculate absolute magnitudes using the distance modulus
%$(m-M)_0=19.90\pm0.06$ that assumes [Fe/H]$=-2.1$ and extinction
%$E(B-V)=0.01\pm0.02$ \citep{lee03}.  The Sextans BHB luminosity function
%appears narrower than our field BHB luminosity functions; a K-S test gives
%a 9\% likelihood for the BHB luminosity functions to be drawn from the
%same distribution.  

\section{CONCLUSIONS}

        We extend the Century Survey Galactic Halo Project based on a new
175 deg$^2$ spectroscopic survey for BHB stars. We make use of the 2MASS
and SDSS photometric catalogs, and show that the 2MASS and SDSS
color-selection is 38\% and 50\% efficient, respectively, for BHB stars.
The 2MASS selection for BHB stars is 65\% {\it complete} \citep{brown04},
but is likely to be worse in the magnitude range $15 < J_0 <15.5$ because
of large photometric errors scattering BHB stars out of the narrow color
selection range. The SDSS completeness for BHB stars is also magnitude
dependent and appears to drop to 50\% in the magnitude range $15 < g'_0
<15.5$ because of saturation problems.

	We analyze the global properties of the original Century Survey,
the 2MASS-selected, and SDSS-selected BHB stars, and find them consistent
with a predominantly halo population.  The median metallicity of the BHB
stars is [Fe/H]=$-1.7$.  K-S tests indicates that the BHB samples share a
common metallicity distribution.  The velocity dispersion of the BHB stars
is 108 km s$^{-1}$.  The mean Galactic rotation of the BHB stars
$3<|z|<15$ kpc is $-4\pm30$ km s$^{-1}$.  Our samples also include a
likely run-away B7 star 6 kpc below the Galactic plane.

	The luminosity of a BHB star is primarily temperature (color)
dependent.  The shape of the $M_V$-color relation is due to the physics of
BHB stars.  We show that the shape of the \citet{preston91}
observationally-derived $M_V$-color relation corresponds to the
\citet{demarque00} theoretical BHB models and to the \citet{kenyon95}
bolometric corrections.  We derive luminosities to our field BHB stars
under the assumption that the $M_V$-color relation is instrinsic to stars
on the BHB.

	The $M_V$-color and $M_V$-metallicity relations impose selection
biases on a flux-limited survey.  A flux-limited survey samples hot BHB
stars with (\bv)$_0=-0.1$ to 64\% of the depth for BHB stars with
(\bv)$_0=+0.1$.  Similarly, flux-limited survey samples metal-rich BHB
stars with [Fe/H]$=-1$ to 82\% of the depth for metal-poor BHB stars with
[Fe/H]$=-3$.

	We calculate the luminosity function for our field BHB star
samples using the maximum-likelihood method of \citet{eep}, a technique
that is non-parametric and unbiased by density inhomogeneities.  The
luminosity function for field BHB stars is characterized by a steep rise
at bright luminosities, a peak between $0.8 < M_V < 1.0$, and a tail at
faint luminosities.  We show that the luminosity function is not
determined by the shape of the $M_V$-color relation, but rather the way
this relation is populated.  We compare our luminosity functions with the
BHB luminosity functions derived from sixteen different globular clusters.  
K-S tests indicate that globular clusters with BHBs, but not globular
clusters with significant extended BHBs, have similar distributions of 
BHB star luminosities as our field BHB star samples.

        We plan to analyze our samples of BHB stars for velocity and
spatial sub-structure. Knowing the global properties and luminosity
function of the BHB stars is an important step in this analysis.
Furthermore, knowing the 2MASS and SDSS color-selection efficiencies and
completenesses for BHB stars are guides our continuing observations. The
eventual goal of our Galactic Halo Project is to identify star streams in
the halo and thus to test the hierarchical picture for galaxy formation.

\acknowledgements

        We thank Perry Berlind and Mike Calkins for their dedicated
observing at the Whipple 1.5 m Tillinghast telescope.  We thank P.\
Demarque and Y.\ Lee for correspondance concerning the BHB $M_V$-color
relation.  We thank the referee for a careful and thoughtful report.  
This project makes use of NASA's Astrophysics Data System Bibliographic
Services. This project makes use of data products from the Two Micron All
Sky Survey, which is a joint project of the University of Massachusetts
and the Infrared Processing and Analysis Center/Caltech, funded by NASA
and the NSF. This project makes use of data products from the Sloan
Digital Sky Survey, which is managed by the Astrophysical Research
Consortium for the Participating Institutions. Funding for SDSS has been
provided by the Sloan foundation, the Participating Institutions, NASA,
NSF, the U.S.\ Dept.\ of Energy, the Japanese Monbukagakusho, and the Max
Planck Society. This work was supported by W.\ Brown's CfA Fellowship.
T.C.B. acknowledges partial support of this work from grants AST 04-06784,
and PHY 02-16783, Physics Frontier Centers/JINA: Joint Institute for
Nuclear Astrophysics, awarded by the NSF.

% \clearpage

\appendix
\section{DATA TABLES}

	Tables \ref{tab:dat1} and \ref{tab:dat2} list the photometric
and spectroscopic measurements for the 2MASS-selected and SDSS-selected
samples.  The tables contain 257 entries and include every 2MASS- and
SDSS-selected object except for the 27 G-type stars in the SDSS-selected
sample.  The SDSS-selected G-types have erroneous photometry, likely due
to saturation problems in the SDSS.  Tables \ref{tab:dat1} and
\ref{tab:dat2} are presented in their entirety in the electronic edition
of the Astronomical Journal.  A portion of the tables are shown here for
guidance regarding their format and content.

% TABLES

% NOTE:  FULL LATEX-FORMATTED DATA TABLE IS tb2.tex
\begin{deluxetable}{lrrccccc}		% TABLE 2
% \tabletypesize \small
% \tablewidth{6.0in}
\tablecaption{PHOTOMETRY\label{tab:dat1}}
\tablecolumns{8}
\tablehead{
	\colhead{} & \colhead{} & \colhead{} &
	\colhead{$J_0$} & \colhead{$g'_0$} &
	\colhead{$E(\bv)$} & \colhead{$BV0$} & \colhead{} \\
        \colhead{ID} &
        \colhead{$\alpha_{\rm J2000}$} &
        \colhead{$\delta_{\rm J2000}$} &
        \colhead{(mag)} & \colhead{(mag)} &
        \colhead{(mag)} & \colhead{(mag)} & \colhead{BHB} \\
        \colhead{(1)} & \colhead{(2)} & \colhead{(3)} &
        \colhead{(4)} & \colhead{(5)} &
        \colhead{(6)} & \colhead{(7)} & \colhead{(8)} \\
}
	\startdata
CHSS 1598 &  3:41:13.2 &  0:48:37 &         \nodata & $15.27\pm0.013$ & 0.09 & $      -0.03$ & 0 \\
CHSS 1599 &  3:43:57.6 &  0:08:57 &         \nodata & $15.11\pm0.020$ & 0.09 & $\phm{-}0.15$ & 0 \\
CHSS 1600 & 23:00:20.9 & -0:17:10 & $14.22\pm0.026$ &         \nodata & 0.05 & $\phm{-}0.16$ & 0 \\
CHSS 1601 & 23:02:10.7 & -1:01:10 & $14.65\pm0.038$ &         \nodata & 0.05 & $      -0.01$ & 1 \\
CHSS 1602 & 23:03:58.3 & -1:08:12 & $13.59\pm0.030$ &         \nodata & 0.04 & $      -0.02$ & 1 \\
	\enddata 
  \tablecomments{Table \ref{tab:dat1} is presented in its entirety
in the electronic edition of the Astronomical Journal.  A portion is shown
here for guidance and content.}
 \end{deluxetable}

% NOTE:  FULL LATEX-FORMATTED DATA TABLE IS tb3.tex
\begin{deluxetable}{ccccccccccc}	% DATA TABLE 3
%\tabletypesize \small
%\tablewidth{6.0in}
\tablecaption{SPECTROSCOPIC AND STELLAR PARAMETERS\label{tab:dat2}}
\tablecolumns{11}
\tablehead{
	\colhead{} & \colhead{} & \colhead{} & \colhead{} &
	\colhead{$v_{radial}$} & \colhead{} &
	\colhead{$T_{eff}$} & \colhead{$\log{g}$} &
	\colhead{} & \colhead{Dist} & \colhead{$M_V$} \\
	\colhead{ID} & \colhead{KP} & \colhead{HP2} & \colhead{GP} &
	\colhead{(km s$^{-1}$)} & \colhead{Type} &
	\colhead{(K)} & \colhead{(cm s$^{-2}$)} &
	\colhead{[Fe/H]} & \colhead{(kpc)} & \colhead{(mag)} \\
	\colhead{(1)} & \colhead{(2)} & \colhead{(3)} & \colhead{(4)} &
	\colhead{(5)} & \colhead{(6)} & 
	\colhead{(7)} & \colhead{(8)} & 
	\colhead{(9)} & \colhead{(10)} & \colhead{(11)} \\
}
	\startdata
CHSS 1598 &    1.97 &        11.00  &    0.89 & $\phm{-1}41.0\pm27.6$ & $22.0\pm1.0$ &  8413 & 4.99 & $      -0.17$ & $\phm{1}5.73$ & $\phm{-}1.28$ \\
CHSS 1599 &    1.72 &        10.11  &    1.38 & $\phm{ }-13.9\pm26.1$ & $31.6\pm2.6$ &  8243 & 4.99 & $      -0.53$ & $\phm{1}1.93$ & $\phm{-}3.60$ \\
CHSS 1600 &    1.90 &        10.07  &    0.81 & $\phm{ }-44.4\pm11.0$ & $21.2\pm1.0$ &  8208 & 4.99 & $      -0.29$ & $\phm{1}2.75$ & $\phm{-}2.09$ \\
CHSS 1601 &    0.47 &        10.91  &    0.05 & $\phm{ }-92.4\pm10.0$ & $21.6\pm1.2$ &  9111 & 3.50 & $      -1.77$ & $\phm{1}4.67$ & $\phm{-}1.23$ \\
CHSS 1602 &    0.31 &        10.91  &    0.27 & $\phm{-11}9.9\pm 9.8$ & $21.6\pm1.2$ &  9095 & 3.50 & $      -1.49$ & $\phm{1}2.65$ & $\phm{-}1.33$ \\
	\enddata	
  \tablecomments{Table \ref{tab:dat2} is presented in its entirety
in the electronic edition of the Astronomical Journal.  A portion is shown
here for guidance and content.}
 \end{deluxetable}

	Table \ref{tab:dat1} summarizes the photometry. Column (1) is our
identifier.  The designation CHSS stands for Century Halo Star Survey and
is chosen to be unique from previous surveys.  Column (2) is the J2000
right ascension in hours, minutes, and seconds.  Column (3) is the J2000
declination in degrees, arcminutes, and arcseconds.  Column (4) is the
2MASS extinction-corrected $J_0$ magnitude for the 2MASS-selected stars.  
Column (5) is the SDSS extinction-corrected $g'_0$ magnitude for the
SDSS-selected stars.  Column (6) is the $E(B-V)$ reddening value from
\citet{schlegel98}.  Column (7) is the $BV0$ color predicted from 2MASS or
SDSS photometry and Balmer line strengths \citep{brown03}.  Column (8) is
the BHB classification:  1 if the star is BHB, 0 if it is not.

	Table \ref{tab:dat2} summarizes the spectroscopic and stellar
parameters.  Column (1) is our identifier.  Column (2) is the KP (Ca
{\small II}) index.  Column (3) is the HP2 (H$\delta$) index.  Column (4)  
is the GP (G-band) index.  Column (5) is the heliocentric radial velocity
in km s$^{-1}$.  Column (6) is the spectral type, where B0=10, A0=20,
F0=30, and so forth.  Column (7) is the effective temperature in K.  
Column (8) is the log base 10 of the surface gravity in cm s$^{-2}$.  
Column (9) is the metallicity given as the logarithmic [Fe/H] ratio
relative to the Sun.  Column (10) is the estimated distance in kpc.  
Column (11) is the absolute $M_V$ magnitude corrected for reddening, given
the estimated distance.

	% REFERENCES 
\clearpage 
% \bibliographystyle{apj} 
% \bibliography{../RefHS}

\end{document}